\documentclass[fleqn,usenatbib]{mnras}

\usepackage[T1]{fontenc}

\DeclareRobustCommand{\VAN}[3]{#2}
\let\VANthebibliography\thebibliography
\def\thebibliography{\DeclareRobustCommand{\VAN}[3]{##3}\VANthebibliography}

\usepackage{graphicx}	
\usepackage{amsmath}	

\usepackage{newtxtext,newtxmath}

\title[Intensity interferometry in photon counting mode]{Stellar intensity interferometry of Vega in photon counting mode}

\author[L. Zampieri et al.]{Luca Zampieri$^{1}$\thanks{E-mail: luca.zampieri@inaf.it}, Giampiero Naletto$^{2,1}$, Aleksandr Burtovoi$^{3,1}$, Michele Fiori$^{2,1}$ and Cesare Barbieri$^{2,1}$
\\
$^{1}$INAF - Astronomical Observatory of Padova, Vicolo dell'Osservatorio 5, 35122, Padova, Italy\\
$^{2}$Department of Physics and Astronomy, University of Padova, Via F. Marzolo 8, 35131, Padova, Italy\\
$^{3}$Centre of Studies and Activities for Space (CISAS) ``G. Colombo'', University of Padova, Via Venezia 15, 35131 Padova, Italy
}

\date{Accepted XXX. Received YYY; in original form ZZZ}

\pubyear{2021}

\begin{document}
\label{firstpage}
\pagerange{\pageref{firstpage}--\pageref{lastpage}}
\maketitle

\begin{abstract}
Stellar Intensity Interferometry is a technique based on the measurement of the second order spatial correlation of the light emitted from a star. The physical information provided by these measurements is the angular size and structure of the emitting source. A worldwide effort is presently under way to implement stellar intensity interferometry on telescopes separated by long baselines and on future arrays of Cherenkov telescopes.
We describe an experiment of this type, realized at the Asiago Observatory (Italy), in which we performed for the first time measurements of the correlation counting photon coincidences in post-processing by means of a single photon software correlator and exploiting entirely the quantum properties of the light emitted from a star. We successfully detected the temporal correlation of Vega at zero baseline and performed a measurement of the correlation on a projected baseline of $\sim$2 km. The average discrete degree of coherence at zero baseline for Vega is $< g^{(2)} > \, = 1.0034 \pm 0.0008$, providing a detection with a signal-to-noise ratio $S/N \gtrsim 4$. No correlation is detected over the km baseline. The measurements are consistent with the expected degree of spatial coherence for a source with the 3.3 mas angular diameter of Vega. The experience gained with the Asiago experiment will serve for future implementations of stellar intensity interferometry on long-baseline arrays of Cherenkov telescopes.
\end{abstract}




\section{Introduction}
\label{sect:intro}

Ordinary (phase) interferometry is widely used in radio Astronomy to measure the size of radio sources and deals with the first order spatial correlation of the radiation emitted from a source. Conversely, intensity interferometry exploits the second order spatial correlation of light \citep{1963PhRv..130.2529G}. A pioneering astronomical experiment of intensity interferometry using the wave nature of light and aiming at measuring stellar radii was performed from the '50s through the '70s of the last century by Hanbury Brown and Twiss \citep{1956Natur.178.1046H,1957RSPSA.242..300B,1958RSPSA.248..199B,1974MNRAS.167..121H,1974iiia.book.....H}. They measured the cross-correlation of the intensity fluctuations of the star signals collected with two photomultipliers at the foci of two 6.5 m telescopes separated by a baseline up to 180 meters. The experiment led to the direct interferometric measurement of the radii of 32 single stars of O-F spectral type \citep{1974MNRAS.167..121H}, greatly improving the scant sample of measurements of K-M giants/supergiants previously obtained with the Michelson's phase interferometer \citep{1931ErNW...10...84P}.

After some preparatory experimental activities carried out by some groups \citep{2016SPIE.9907E..0NZ,2016MNRAS.457.4291T,2018SPIE10701E..0WM,2018SPIE10701E..0XW}, new successful stellar intensity interferometry (SII) measurements a l\`a Hanbury Brown and Twiss have recently been realized using the particle nature of light and modern fast single-photon counters. The first intensity correlation measured with starlight from conventional optical telescopes since the historical experiments of Hanbury Brown and Twiss has been performed by \cite{2017MNRAS.472.4126G}, and more recently by \cite{2020MNRAS.494..218R}. The renewed interest for stellar intensity interferometry sparkled from the planned realization of extended arrays of Cherenkov telescopes for Very High Energy Astronomy. They will have both large collecting areas and a large number of baselines, from $\sim$100 meters up to a $\sim$1 kilometer, thus indirectly providing a suitable infrastructure for performing SII measurements and visible image reconstruction with an unprecedented spatial resolution \citep{2006ApJ...649..399L,2012MNRAS.419..172N,2012MNRAS.424.1006N,2013APh....43..331D,2013MNRAS.430.3187R,2019BAAS...51c.275K}. These measurements will allow us to do science that has not yet been possible before. Recently, the capability of performing SII measurements and the potential of the technique with the MAGIC and VERITAS Cherenkov telescopes has been convincingly demonstrated by \cite{2020MNRAS.491.1540A} and \cite{2020NatAs...4.1164A}, respectively.

In this context, in 2015 and 2016 we started the first experimental runs of the Asiago Intensity Interferometer, using our fast photon counters Aqueye+ and Iqueye \citep{2016SPIE.9907E..0NZ,2016SPIE.9980E..0GN}. The instrumentation allows us both to detect the correlation of the signal at essentially zero baseline (exploiting the instrument internal sub-apertures) and to perform measurements on long separations, 
thus demonstrating the feasibility of km-baseline-long photon counting SII.

Between 2017 and 2018 we devoted several runs to reaching an adequate control of the systematic errors and to implementing an efficient approach for the single-photon data reduction and analysis. The final instrumental set-up and data analysis technique, that led to a successfull implementation in 2019, are reported here.

Because of the small collecting area of the telescopes, we focussed on a bright target. We selected Vega, the second brightest star in the Northern hemisphere ($V=0.03$, \citealt{2002yCat.2237....0D}; A0Va spectral type \citealt{2003AJ....126.2048G}). Its angular diameter is $\simeq 3.3$ mas ($3.28 \pm 0.01$ mas, \citealt{2001ApJ...559.1147C}; 3.324 mas, \citealt{2012ApJ...761L...3M}). Despite being one of the brightest stars in the sky, Vega quite recently revealed new and unexpected properties. Optical interferometric observations showed that the star has the asymmetric brightness distribution of the slightly offset polar axis of a star rotating at 93\% of its breakup speed \citep{2006Natur.440..896P}.

The plan of the paper is the following. In Section~\ref{sect:instrumentation} we describe our instrumental interferometric setup. In Section~\ref{sect:analysis} we report the post-processing single-photon data analysis technique adopted for calculating the degree of correlation, and in Section~\ref{sect:observations} we list the Aqueye+ anf IFI+Iqueye observations of Vega carried out in July-August and November 2019. In Section~\ref{sect:instr-systematics} we discuss the instrument systematics and the final adopted calibration of our measurements. Finally, in Section~\ref{sect:results} we show the results of our analysis and in Section~\ref{sect:discussion} we shortly discuss the possible future implementations of our photon counting approach for measurements of stellar intensity interferometry.

\begin{table}
\caption{Coordinates, distance and baseline of the Galileo (T122) and Copernicus (T182) telescopes in Asiago. Coordinates refer to the intersections of the hour angle and declination axes.}
\label{tab:coordinates}
\centering
\begin{tabular}{l r r}
\hline\hline
\multicolumn{3}{c}{Geographic and Cartesian geocentric coordinates}	\\
\hline
                                        &     Geographic         &     Cartesian (m)  \\
\hline
T122                                    &  11 31 35.14 E  (Long) &    4360008.6 (X) \\
                                        &  45 51 59.22 N  (Lat)  &     889148.3 (Y) \\
                                        &     1094.6  m   (Elev$^1$) &    4555709.2 (Z) \\
\hline
T182                                    &  11 34 08.397 E (Long) &    4360935.4 (X) \\
                                        &  45 50 54.894 N (Lat)  &     892712.8 (Y) \\
                                        &     1376.2 m    (Elev$^1$) &    4554527.8 (Z) \\
\hline
\multicolumn{3}{c}{Distance T182-T122 (m)}	\\
\hline
\multicolumn{3}{c}{281.6 (Elev$^1$)}    \\
\multicolumn{3}{c}{3867.8 (Total)}  \\
\hline
\multicolumn{3}{c}{Maximum projected baseline T182-T122 (m)}	\\
\hline
\multicolumn{3}{c}{3213.8 (E-W)}    \\
\multicolumn{3}{c}{2133.6 (N-S)}    \\
\multicolumn{3}{c}{3857.6 (Total)}  \\
\hline
\end{tabular}
\begin{minipage}{8.2 cm}
$^1$ Elevation measured from sea level
\end{minipage}
\end{table}

\section{The Asiago stellar intensity interferometer}
\label{sect:instrumentation}

Aqueye+ and Iqueye\footnote{http://web.oapd.inaf.it/zampieri/aqueye-iqueye/index.html} are narrow field-of-view photon counting photometers with sub-nanosecond absolute time accuracy \citep{2009JMOp...56..261B,2009A&A...508..531N,2013SPIE.8875E..0DN,2015SPIE.9504E..0CZ}. Their main features are: a field of view of a few arcsec, a 4-split pupil optical design achieved using a pyramidal mirror, Single Photon Avalanche Diode (SPAD) detectors with tens of ps time resolution, an acquisition system capable of sub-ns time tagging accuracy with respect to UTC. The 4-split pupil optical design causes the incoming beam to be divided in four sub-apertures that are focussed on four independent SPADs. The four sub-apertures allow us to perform a cross-correlation of the signal also at zero baseline \citep{2016SPIE.9907E..0NZ,2016SPIE.9980E..0GN}\footnote{See also \cite{2010SPIE.7702E..0MC} for a preliminary measurement at zero baseline performed at the New Technology Telescope, in Chile.}, which is crucial to calibrate the degree of coherence. In the following we will refer to the sub-apertures of Aqueye+ and Iqueye with $A$, $B$, $C$ and $D$, where $A-C$ and $B-D$ represent the two baselines with face-to-face detectors.

The main observing facilities in Asiago (Italy), the 1.22 m Galileo telescope and the 1.82 m Copernicus telescope, are located in the resorts of Pennar and Cima Ekar, almost 4 km apart. Equipped with Aqueye+ and Iqueye, the two telescopes are well suited to realize a photon counting km-baseline intensity interferometer (see Table~\ref{tab:coordinates}). Aqueye+ is directly mounted at the Copernicus telescope, while Iqueye is fiber-coupled with the Galileo telescope by means of a dedicated instrument, the Iqueye Fiber Interface (IFI; \citealt{2019CoSka..49...85Z}).


The measurements were performed using two different sets of filters, an H$\alpha$ filter plus a $\times$10 neutral density filter (ND1) and a narrow band interferometric filter (hereafter referred to as II filter; see Table~\ref{tab:filters}). All the acquired data are stored for the post-processing analysis.

\section{Discrete degree of coherence and photon-counting software correlator}
\label{sect:analysis}

The main observable for SII is the second order (discrete) degree of coherence $g^{(2)}(\tau,d)$ of a star, that measures the degree of correlation of its light and depends on the telescopes/sub-apertures separation $d$ and the relative delay $\tau$ between them. We calculate $g^{(2)}(\tau,d)$ in post-processing using the expression (e.g. \citealt{2016SPIE.9980E..0GN,2016SPIE.9907E..0NZ}):
\begin{equation}
  g^{(2)}(\tau,d) = \frac{N_{XY} N}{N_X N_Y} \, ,
  \label{eq:eq1}
\end{equation}
where 
$N_X$ and $N_Y$ are the number of photons detected on the sub-apertures $X$ and $Y$ (of the same telescope or of two different telescopes) in a time interval $T_s$, $N_{XY}$ is the number of simultaneous detections (coincidences) in both sub-apertures in a small time bin $dt$, and $N=T_s/dt$ is the total number of bins in time $T_s$. The major contribution to $N_{XY}$ comes from random uncorrelated coincidences. The signal is a tiny excess of coincidences related to the quantum nature of light (bosons giving a joint detection probability greater than that for two independent events).

\begin{table}
\caption{Filters used for the 2019 Aqueye+ and IFI+Iqueye observations of Vega.}
\label{tab:filters}
\centering
\begin{tabular}{l c c c}
\hline\hline
\multicolumn{4}{c}{Filters}	\\
\hline
Filter           & $\lambda_c^a$ (nm) & FWHM$^b$ (nm) & peak transmission (\%) \\
\hline
H$\alpha$+ND1	 &   656.7        &  3   & 62 \\
II               &   510.5        & 0.3  & 35 \\
\hline
\end{tabular}
\begin{minipage}{8.2 cm}
$^a$ Central wavelength \\
$^b$ Full width half maximum 
\end{minipage}
\end{table}

A dedicated software package (Aqueye+/Iqueye software correlator, written in Linux bash shell, Fortran, Python) was developed for determining the number of coincidences $N_{XY}$ and the degree of coherence $g^{(2)}$ of our event lists. The correlation is entirely performed in post-processing using the following procedure:
\begin{enumerate}
\item The (non-barycentered) event lists are divided in $N_s$ segments of duration $T_s = 8.64$ s and then searched for coincidences $N_{XY}$ in time bins of duration $dt \simeq 400$ ps. The choice of the time bin is discussed below.
\item The degree of coherence $g^{(2)}$ is then calculated in each segment using equation~(\ref{eq:eq1}) and the values are then averaged out over the various segments of an acquisition. An additional average is performed over all possible combinations $X-Y$ of the sub-apertures. To calculate the temporal correlation a delay $\tau$ is added or subtracted to the photons of one sub-aperture and the calculation is then repeated. The delay is taken in steps of $\simeq$200 ps in the interval between -20.5 ns and +20.5 ns. The average value of $g^{(2)}(\tau,d)$ for the $k$-th acquisition is then:
\begin{equation}
  g^{(2)}_k (\tau,d) = \frac{1}{N_b N_s} \sum_{j=1}^{N_b} \sum_{i=1}^{N_s} g^{(2)}_{ij} \, ,
  \label{eq:eq2}
\end{equation}
where $i$ is the summation index over the $N_s$ time segments and $j$ that over the $N_b$ sub-apertures.
\item The final calibrated value of $g^{(2)}(\tau,d)$ is calculated subtracting the measurement averaged over the acquisitions with the H$\alpha$+ND1 filters from the measurement averaged over the acquisitions with the II filter, i.e.:
\begin{equation}
  < g^{(2)} (\tau,d) > = 1 + <g^{(2)}(\tau,d)>_{II} - <g^{(2)}(\tau)>_{{\rm H}\alpha+{\rm ND1}} \, ,
  \label{eq:eq3}
\end{equation}
where the average is over the acquisitions.
\item If the amount of data is sufficient, the calculation is done independently for each observing night and a final average of $g^{(2)}(\tau,d)$ is then performed using the measurements acquired each night. For the observations reported here, this was possible only for the zero baseline data acquired with Aqueye+.
\end{enumerate}

The search for coincidences in step (1) is performed after binning the event lists. Therefore, two photons are considered coincident in time bin $dt$ if their arrival time is within the bounds of the time bin. If two photons do not fall inside the same bin, they are not considered coincident, even if the difference of their arrival times is smaller than $dt$. The algorithm is optimized to record only the time bins in which a photon detection actually occurs, discarding all the others. As discussed below, the exquisite accuracy of our timing allows us to push the sampling time $dt$ to the limit and we eventually selected $dt \simeq 400$ ps.

The expected theoretical signal-to-noise ratio of a measurement of $g^{(2)}(0,d)$ in a time interval $T$ and with a sampling time $dt$ is (e.g. \citealt{2016SPIE.9980E..0GN,2016SPIE.9907E..0NZ}):
\begin{equation}
S/N = n (\lambda/c) (\lambda/\Delta \lambda) \alpha |\gamma(0,d)|^2 [T/(2dt)]^{1/2} \, , 
\label{eq:sn}
\end{equation}
where $n$ is the geometric average of the source count rate over two sub-apertures (or telescopes) in photons per second in the optical bandpass $\Delta \lambda$, $\lambda$ is the central wavelength of the bandpass, $\alpha$ the detector efficiency, and $|\gamma(0,d)|^2$ the square modulus of the degree of coherence at zero delay. Plugging in the values for our measurement/instrumental set-up ($n \sim 1$ Mc/s, $\lambda = 510.5$ nm, $\Delta\lambda = 0.3$ nm, $\alpha = 0.5$, $T \sim 30$ minutes, $dt \sim 400$ ps) and assuming full correlation at zero baseline ($|\gamma(0,d)|^2 = 1$), we obtain $S/N \sim 4$ at $d=0$. Thus, despite the short acquisition time we expected to be able to obtain a significant measurement of $g^{(2)}$.

\subsection{Choice of the time bin $dt$}
\label{sect:analysis-dt}

An important technical aspect of the measurement is the choice of the time bin $dt$. The very high time accuracy of the acquisition chain of Aqueye+ and Iqueye allows us to push this parameter at the limit and gain on the signal to noise ratio of the measurement. For the zero baseline measurement we are limited only by the relative time accuracy among the sub-apertures, which is $\simeq$100 ps. The most stringent constraints come from the absolute time accuracy when correlating the data from the two telescopes. The error induced by the correction for the light travel time delay between them is typically $\simeq$200 ps. The time dispersion induced by the multimode optical fiber injecting the star light into Iqueye is $\lesssim$250 ps (see eq.~[6] in \citealt{2016SPIE.9907E..0NZ})\footnote{The focal length of the lens injecting light into the optical fiber was changed and is now 100 mm.}. Therefore, the final choice is dictated mainly by the absolute accuracy of the photon arrival times with respect to UTC, which is $\leq$1.5-2 ns for the typical clock drift and acquisition length of the observations used here. With a time bin $dt \simeq 400$ ps correlated photons will then spread over 4-5 adjacent time bins, leading to a decrease of the signal-to-noise ratio by a factor $\simeq$2 (eq.~[\ref{eq:sn}]). This would not allow us to achieve a significant detection of any potential correlation between the two telescopes, but is sufficient to exclude that a correlation exists, as shown in Section~\ref{sect:results}. On the other hand, for the measurement at zero baseline, a time bin $dt \simeq 400$ ps allows us to achieve a signal to noise ratio adequate for a detection (eq.~[\ref{eq:sn}]). The actual time bin was set to 16 times the resolution of the time-to-digital-converter (24.2 ps) and is then $dt \simeq 387$ ps.

\section{Observations and data analysis}
\label{sect:observations}

We report on the results of two runs devoted to intensity interferometry observations of Vega in Asiago, the first performed on 2019 July 31-August 1 and the second on 2019 November 25-28. The log of the acquisitions is shown in Table~\ref{tab:observations}. During the 2019 Jul-Aug run we specifically aimed at detecting the temporal correlation of the light from Vega using the sub-apertures of Aqueye+, and hence no IFI+Iqueye observations were carried out. Simultaneous acquisitions with both instruments were done during the November run.

We retained only the acquisitions for which sky conditions were good (no significant veils or clouds). A total of 38 minutes of useful data with both the H$\alpha$ and the II filters were acquired with Aqueye+ on Jul 31-Aug 1, 2019. The total duration of the simultaneous Aqueye+/IFI+Iqueye acquisitions (Nov 25+28) was 28 minutes with the H$\alpha$ filter and 39 minutes with the II filter. The average count rate measured with Aqueye+/IFI+Iqueye was $\sim$1.9/0.1 Mc/s in the H$\alpha$ filter and $\sim$2.7/0.2 Mc/s in the II filter.

The preliminary reduction of the data was performed using a dedicated software \citep{2015SPIE.9504E..0CZ}. The whole acquisition and reduction chain ensures an absolute time accuracy of $\sim$0.5 ns with respect to UTC and a relative accuracy in a single acquisition of $\approx$100 ps \citep{2009A&A...508..531N}. The intensity interferometry data analysis was done in post-processing as described in Section~\ref{sect:analysis}.

\begin{table*}
\scriptsize
\centering
\caption{Log of the 2019 July 31-Aug 1 and 2019 November 25-28 observations of Vega taken with Aqueye+ at the Copernicus telescope and IFI+Iqueye at the Galileo telescope in Asiago.}
\label{tab:observations}
\begin{tabular}{llrllrllr}
\hline
\hline
\multicolumn{3}{c}{31 Jul-1 Aug 2019$^1$} & \multicolumn{3}{c}{25 Nov 2019$^2$} & \multicolumn{3}{c}{28 Nov 2019$^3$}	\\
\hline
\multicolumn{1}{l}{Observation ID} & \multicolumn{1}{l}{Filter} & \multicolumn{1}{r}{Duration$^4$ (s)} & \multicolumn{1}{l}{Observation ID$^5$} & \multicolumn{1}{l}{Filter} & \multicolumn{1}{r}{Duration$^{6}$ (s)} & \multicolumn{1}{l}{Observation ID$^5$} & \multicolumn{1}{l}{Filter} & \multicolumn{1}{r}{Duration$^{6}$ (s)} \\
\hline                                                                
20190731-215000   & II               &     12    &   20191125-191218        & II               &    60        &   20191128-192237         & II              &    60         \\
20190731-215642   & II               &     61    &   20191125-191351        & II               &    60        &   20191128-192406         & II              &    60         \\
20190731-220359   & II               &     63    &   20191125-191522        & II               &    60        &   20191128-193750 (A+I)   & II              &    60 (52)    \\
20190731-221812   & H$\alpha$+ND1    &     60    &   20191125-191723        & II               &    60        &   20191128-193926 (A+I)   & II              &    60 (52)    \\
20190731-222450   & H$\alpha$+ND1    &    121    &   20191125-191856        & II               &    60        &   20191128-194157         & II              &    60         \\
20190731-223128   & H$\alpha$+ND1    &     89    &   20191125-192030        & II               &    60        &   20191128-194406         & II              &    60         \\
20190731-224153   & II               &    155    &   20191125-192208        & II               &    60        &   20191128-194542 (A+I)   & II              &    60 (43)    \\
20190731-224839   & II               &     32    &   20191125-192340        & II               &    60        &   20191128-194716 (A+I)   & II              &    60 (43)    \\
20190731-225518   & II               &     32    &   20191125-192521        & II               &    60        &   20191128-194853 (A+I)   & II              &    60 (52)    \\
20190731-230447   & H$\alpha$+ND1    &     60    &   20191125-192652        & II               &    60        &   20191128-195028 (A+I)   & II              &    60 (52)    \\
20190731-231122   & H$\alpha$+ND1    &     63    &   20191125-192827        & II               &    60        &   20191128-195203 (A+I)   & II              &    60 (52)    \\
20190731-231756   & H$\alpha$+ND1    &    127    &   20191125-192957        & II               &    60        &   20191128-195339 (A+I)   & II              &    60 (52)    \\
20190731-233324   & II               &     22    &   20191125-193130        & II               &    60        &   20191128-195518 (A+I)   & II              &    60 (52)    \\
20190731-234119   & II               &     62    &   20191125-193312        & II               &    60        &   20191128-195702 (A+I)   & II              &    60 (52)    \\
20190731-234757   & II               &     61    &   20191125-193552        & H$\alpha$+ND1    &    60        &   20191128-195839 (A+I)   & II              &    60 (52)    \\
20190731-235904   & H$\alpha$+ND1    &     14    &   20191125-193724        & H$\alpha$+ND1    &    60        &   20191128-200019 (A+I)   & II              &    60 (52)    \\
20190801-000547   & H$\alpha$+ND1    &     63    &   20191125-194412        & H$\alpha$+ND1    &    60        &   20191128-200153 (A+I)   & II              &    60 (52)    \\
20190801-001322   & H$\alpha$+ND1    &     96    &   20191125-194543        & H$\alpha$+ND1    &    60        &   20191128-200326 (A+I)   & II              &    60 (52)    \\
20190801-002259   & II               &     14    &   20191125-194718        & H$\alpha$+ND1    &    60        &   20191128-200457 (A+I)   & II              &    60 (52)    \\
20190801-002940   & II               &     63    &   20191125-194853        & H$\alpha$+ND1    &    60        &   20191128-200840         & H$\alpha$+ND1   &    60         \\
20190801-003628   & II               &     63    &   20191125-195026        & H$\alpha$+ND1    &    60        &   20191128-201010         & H$\alpha$+ND1   &    60         \\
20190801-004723   & H$\alpha$+ND1    &    102    &   20191125-195158        & H$\alpha$+ND1    &    60        &   20191128-201141         & H$\alpha$+ND1   &    60         \\
20190801-005407   & H$\alpha$+ND1    &     90    &   20191125-195337        & H$\alpha$+ND1    &    60        &   20191128-201321         & H$\alpha$+ND1   &    60         \\
20190801-010044   & H$\alpha$+ND1    &     61    &   20191125-195640        & H$\alpha$+ND1    &    60        &   20191128-201547 (A+I)   & H$\alpha$+ND1   &    60 (43)    \\
20190801-011050   & II               &     68    &   20191125-195815        & H$\alpha$+ND1    &    60        &   20191128-201742 (A+I)   & H$\alpha$+ND1   &    60 (52)    \\
20190801-011727   & II               &     60    &   20191125-195951        & H$\alpha$+ND1    &    60        &   20191128-201914 (A+I)   & H$\alpha$+ND1   &    60 (43)    \\
20190801-012402   & II               &     58    &   20191125-200125        & H$\alpha$+ND1    &    60        &   20191128-202055 (A+I)   & H$\alpha$+ND1   &    60 (52)    \\
20190801-013512   & H$\alpha$+ND1    &     84    &   20191125-200257        & H$\alpha$+ND1    &    60        &   20191128-202230 (A+I)   & H$\alpha$+ND1   &    60 (52)    \\
20190801-014148   & H$\alpha$+ND1    &     91    &   20191125-200501        & II               &    60        &   20191128-202403 (A+I)   & H$\alpha$+ND1   &    60 (52)    \\
20190801-014824   & H$\alpha$+ND1    &     62    &   20191125-200633        & II               &    60        &   20191128-202541 (A+I)   & H$\alpha$+ND1   &    60 (52)    \\
20190801-015849   & II               &     13    &   20191125-200807        & II               &    60        &   20191128-202714 (A+I)   & H$\alpha$+ND1   &    60 (52)    \\
20190801-020547   & II               &     58    &   20191125-200936        & II               &    60        &   20191128-202900 (A+I)   & H$\alpha$+ND1   &    60 (52)    \\
20190801-021220   & II               &     58    &   20191125-201107        & II               &    60        &   20191128-203034 (A+I)   & H$\alpha$+ND1   &    60 (52)    \\
20190801-022326   & H$\alpha$+ND1    &     24    &   20191125-201251        & II               &    60        &   20191128-203224 (A+I)   & H$\alpha$+ND1   &    60 (52)    \\
20190801-023002   & H$\alpha$+ND1    &    121    &   20191125-201431        & II               &    60        &   20191128-203406 (A+I)   & H$\alpha$+ND1   &    60 (52)    \\
20190801-023642   & H$\alpha$+ND1    &    119    &   20191125-201606        & II               &    60        &   20191128-203543 (A+I)   & H$\alpha$+ND1   &    60 (52)    \\
20190801-024913   & II               &     86    &   20191125-201738        & II               &    60        &   20191128-203723 (A+I)   & H$\alpha$+ND1   &    60 (43)    \\
20190801-025558   & II               &     59    &   20191125-201911        & II               &    60        &   20191128-203857 (A+I)   & H$\alpha$+ND1   &    60 (52)    \\
20190801-030259   & II               &     30    &   20191125-202044        & II               &    60        &   20191128-204118         & II              &    60         \\
20190801-031310   & H$\alpha$+ND1    &     60    &   20191125-202227        & II               &    60        &   20191128-204250         & II              &    60         \\
20190801-031945   & H$\alpha$+ND1    &     62    &   20191125-202403        & II               &    60        &   20191128-204437         & II              &    60         \\
20190801-032621   & H$\alpha$+ND1    &     58    &   20191125-202552        & II               &    60        &   20191128-204615         & II              &    60         \\
20190801-033728   & II               &    122    &   20191125-202725        & II               &    60        &   20191128-204759         & II              &    60         \\
20190801-034405   & II               &     58    &   20191125-202944 (A+I)  & H$\alpha$+ND1    &    60 (52)   &   20191128-204939 (A+I)   & II              &    60 (52)    \\
20190801-035042   & II               &     89    &   20191125-203305 (A+I)  & H$\alpha$+ND1    &    60 (52)   &   20191128-205122 (A+I)   & II              &    60 (52)    \\
20190801-040026   & H$\alpha$+ND1    &    176    &   20191125-203441 (A+I)  & H$\alpha$+ND1    &    60 (52)   &   20191128-205253 (A+I)   & II              &    60 (52)    \\
20190801-040702   & H$\alpha$+ND1    &     62    &   20191125-203618 (A+I)  & H$\alpha$+ND1    &    60 (52)   &   20191128-205428 (A+I)   & II              &    60 (52)    \\
20190801-041344   & H$\alpha$+ND1    &     93    &   20191125-203757        & H$\alpha$+ND1    &    60        &   20191128-205603 (A+I)   & II              &    60 (52)    \\
                  &                  &           &   20191125-203941 (A+I)  & H$\alpha$+ND1    &    60 (43)   &   20191128-205743 (A+I)   & II              &    60 (52)    \\
                  &                  &           &   20191125-204119 (A+I)  & H$\alpha$+ND1    &    60 (52)   &   20191128-205920 (A+I)   & II              &    60 (52)    \\
                  &                  &           &   20191125-204256 (A+I)  & H$\alpha$+ND1    &    60 (52)   &   20191128-210056 (A+I)   & II              &    60 (52)    \\
                  &                  &           &   20191125-204429 (A+I)  & H$\alpha$+ND1    &    60 (52)   &   20191128-210244 (A+I)   & II              &    60 (52)    \\
                  &                  &           &   20191125-204607 (A+I)  & H$\alpha$+ND1    &    60 (52)   &   20191128-210418 (A+I)   & II              &    60 (52)    \\
                  &                  &           &   20191125-204758 (A+I)  & H$\alpha$+ND1    &    60 (52)   &   20191128-210552 (A+I)   & II              &    60 (52)    \\
                  &                  &           &   20191125-204935 (A+I)  & H$\alpha$+ND1    &    60 (52)   &   20191128-210725 (A+I)   & II              &    60 (52)    \\
                  &                  &           &   20191125-205112 (A+I)  & H$\alpha$+ND1    &    60 (43)   &   20191128-211053 (A+I)   & II              &    60 (43)    \\
                  &                  &           &   20191125-205253 (A+I)  & H$\alpha$+ND1    &    60 (52)   &   20191128-211237 (A+I)   & II              &    60 (52)    \\
                  &                  &           &   20191125-205536        & II               &    60        &                           &                 &               \\
                  &                  &           &   20191125-205703        & II               &    60        &                           &                 &               \\
                  &                  &           &   20191125-205833        & II               &    60        &                           &                 &               \\
                  &                  &           &   20191125-210001        & II               &    60        &                           &                 &               \\
                  &                  &           &   20191125-210133        & II               &    60        &                           &                 &               \\
                  &                  &           &   20191125-210307 (A+I)  & II               &    60 (52)   &                           &                 &               \\
                  &                  &           &   20191125-210446 (A+I)  & II               &    60 (52)   &                           &                 &               \\
                  &                  &           &   20191125-210620 (A+I)  & II               &    60 (52)   &                           &                 &               \\
                  &                  &           &   20191125-210753 (A+I)  & II               &    60 (52)   &                           &                 &               \\
                  &                  &           &   20191125-210931 (A+I)  & II               &    60 (52)   &                           &                 &               \\
                  &                  &           &   20191125-211114 (A+I)  & II               &    60 (52)   &                           &                 &               \\
                  &                  &           &   20191125-211246 (A+I)  & II               &    60 (52)   &                           &                 &               \\
                  &                  &           &   20191125-211422 (A+I)  & II               &    60 (52)   &                           &                 &               \\
                  &                  &           &   20191125-211556 (A+I)  & II               &    60 (52)   &                           &                 &               \\
                  &                  &           &   20191125-211729 (A+I)  & II               &    60 (52)   &                           &                 &               \\
\hline                                                           
\end{tabular}
\begin{minipage}{10.2 cm}
$^1$ Start time (UTC): MJD 58695.826400, stop time (UTC): MJD 58696.093971 --- $^2$ Start time (UTC): MJD 58812.757486, stop time (UTC): MJD 58812.846181 --- $^3$ Start time (UTC): MJD 58815.765716, stop time (UTC): MJD 58815.842789 --- $^4$ Rounded to 1 s --- $^5$ (A+I) identifies simultaneous Aqueye and IFI+Iqueye observations --- $^6$ The number in round brackets is the simultaneous acquisition time used in the analysis, rounded to 1 s
\end{minipage}
\end{table*} 

\section{Instrument systematics}
\label{sect:instr-systematics}

Before presenting the results of our analysis, in this Section we discuss how we reached an adequate control of the instrumentation and the observational strategy after identifying a number of systematic effects in our measurements. These are crucial issues to investigate, especially when performing low signal-to-noise SII measurements in photon counting and post-processing.

\begin{figure}
	\includegraphics[angle=0, width=0.94\columnwidth]{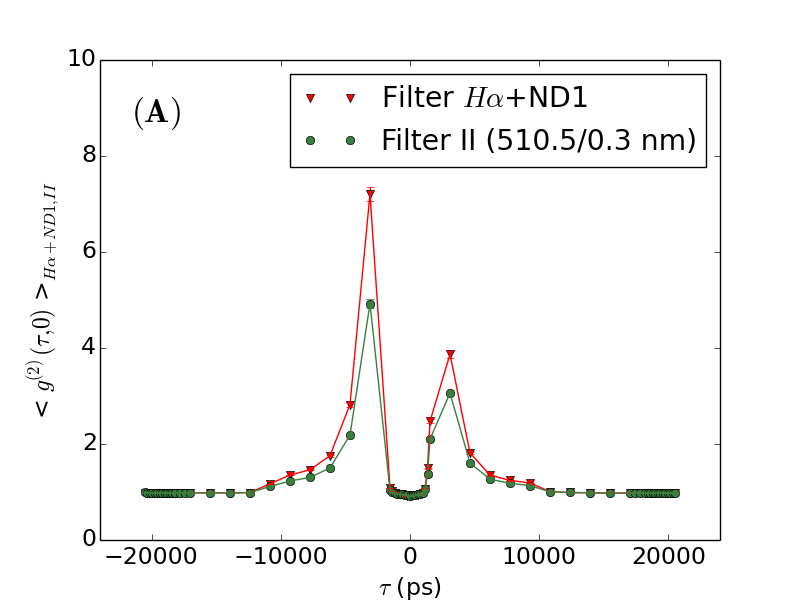}
	\includegraphics[angle=0, width=0.94\columnwidth]{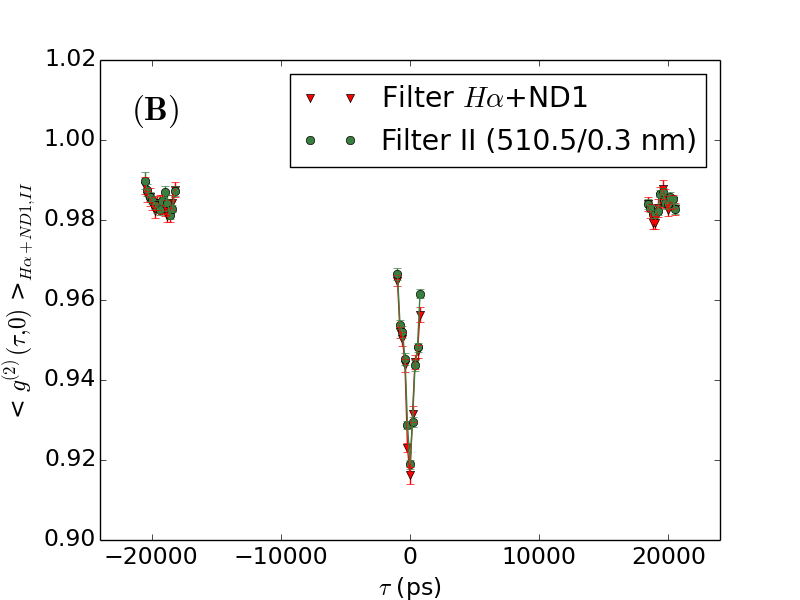}
	\includegraphics[angle=0, width=0.94\columnwidth]{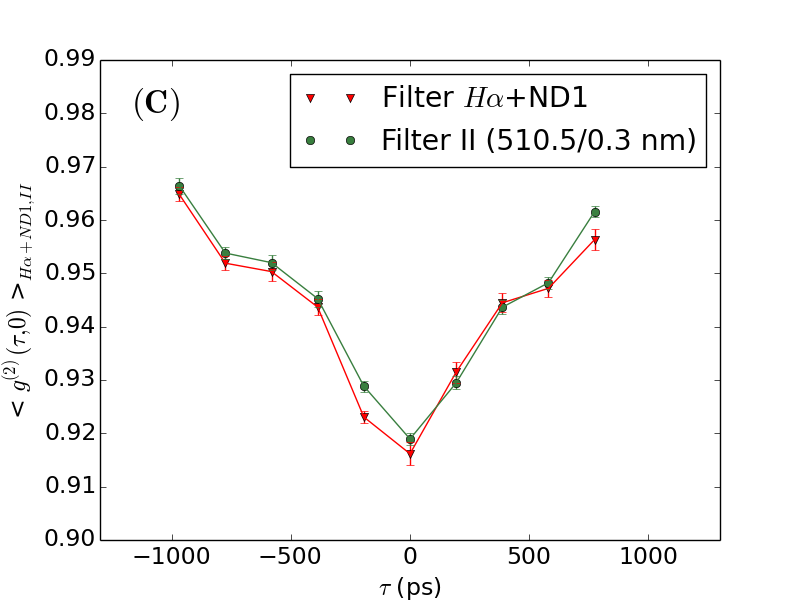}
    \caption{Temporal correlation at zero baseline for the Aqueye+ observations of Vega taken on Nov 28, 2019, calculated with a sampling time $dt \simeq 400$ ps and for the two adopted combinations of filters, H$\alpha$+ND1 and II. The properties of the two sets of filters are reported in Table~\ref{tab:observations}. {\it Panel (A)}: measurements for delays $\tau=[-20.5,20.5]$ ns. {\it Panel (B)}: measurements after removing the intervals $\tau=[-18,-1]$ ns and $\tau=[0.8,18]$ ns. {\it Panel (C)}: measurements for delays $\tau=[-1,0.8]$ ns.}
    \label{fig:g2taufilters}
\end{figure}

Figure~\ref{fig:g2taufilters} shows the temporal correlation at zero baseline for the Aqueye+ observations of Vega taken on Nov 28, 2019. As can be seen in panel A, $g^{(2)}$ exceeds dramatically the expected value of 1 for delays in the intervals $\tau \simeq [-12, -1]$ ns and $\tau \simeq [1, 12]$ ns. This sharp excess is the consequence of a systematic effect.  Aqueye+ and Iqueye are affected by spurious photon coincidences caused by secondary photons emitted when a primary photon hits a SPAD detector (e.g \citealt{2007SPIE.6771E..11R}). Even if we paid particular care in minimizing this effect (by inserting diaphragms and anti-reflective coatings), a tiny fraction of the secondary photons can still be channeled along the optical path back to another SPAD.

We noted that spurious coincidences affect in a significant way the measurement of $g^{(2)}$ on adjacent baselines of the same instrument ($A-B$, $B-C$, $C-D$,$D-A$), while the cross-baselines ($A-C$, $B-D$) are much less affected. This fact is a consequence of the internal structure of Aqueye+ and Iqueye, that channels secondary photons mostly in the direction of the detector in front. Indeed, we found that the most pronounced peaks of $g^{(2)}$ at a delay of a few ns (Figure~\ref{fig:g2taufilters}) are present in the temporal correlation of the cross-baselines. As a consequence, the burst of secondary photons impinging on a detector has a certain probability to produce a spurious coincidence with a real photon that, during the burst interval, hits an adjacent detector.

\begin{table}
\caption{Combination of Aqueye+ sub-apertures used for the measurement of the correlation at zero baseline.}
\label{tab:baselines}
\centering
\begin{tabular}{l c}
\hline
Observing night          & Baselines$^{1}$   \\
\hline
31 Jul-1 Aug 2019	 & $A-B$, $A-C$, $B-C$ \\
25 Nov 2019              & $A-B$, $A-C$, $A-D$, $B-D$, $C-D$ \\
28 Nov 2019              & $A-C$, $A-D$, $B-D$, $C-D$ \\
\hline
\end{tabular}
\begin{minipage}{8.2 cm}
$^1$ $A$, $B$, $C$ and $D$ are the instrument sub-apertures (see text for details).
\end{minipage}
\end{table}

For this reason, in the Jul-Aug 2019 run we opted for a dedicated set-up, inserting an additional filter in the sub-aperture $D$ of Aqueye+, with the purpose of suppressing the
flux of secondary photons and the related background contamination. The test performed that night showed that spurious coincidences affect the photon flux in each channel 
at the level of $\gtrsim$1\%. Inserting an additional filter solved the problem, but it caused a significant suppression of the flux on the sub-aperture D with the additional II filter, that has a peak transmission of only 35\% (see Table~\ref{tab:filters}). For Jul-Aug 2019 we then considered the less noisy measurement with the baselines made only by high counting statistics sub-apertures ($A-B$, $A-C$, $B-C$; Table~\ref{tab:baselines}).

Conversely, in the Nov 2019 run we mounted the two available II filters separately in the two instruments. Therefore, we could not decrease the secondary-photons-induced noise
but, in principle, we could take advantage of the high counting statistics of all sub-apertures for the measurement of the correlation at zero baseline. However, on Nov 25 the baseline $B-C$ and on Nov 28 the baselines $A-B$ and $B-C$ showed an anomalous behaviour and we had to remove them (Table~\ref{tab:baselines}).

The reason behind this anomalous behaviour is related to another important instrumental systematics, the spurious (anti-)correlations between different channels of the front-end electronics that reads the signals from all the detectors at a given telescope. Eventually, most of them turned out to produce rather stable patterns in the measurement of $g^{(2)}$ and we succeeded in removing them by subtracting the measurements performed with two different filters (see below). The residual systematic offset induced by this effect is $\lesssim$0.0005. However, at times, two baselines ($A-B$ and $B-C$) appear to have a variable response on a rather short time scale (up to a few minutes), significantly dependent on ambient conditions (e.g. temperature). Therefore, on these baselines, variations of the average value of $g^{(2)}$ measured in observations taken only a few minutes apart can be anomalously large ($\sim$1\%). In this case the two-filters-approach is not successful in removing the spurious (anti-)correlations patterns. The reason for which this happens is still unclear (maybe an anomalous coupling/variation of the frequency of the jitter of channels $A$ and $B$, or $B$ and $C$) but, when the problem appeared during an observing night, we discarded the corresponding baseline from the final average of the measurements. As mentioned above, this was the case for baseline $B-C$ on Nov 25, and for baselines $A-B$ and $B-C$ on Nov 28.

\subsection{Calibration of the measurements}
\label{sect:instr-calib}

As can be seen in Figure~\ref{fig:g2taufilters} (panel A), the cross-talk effects caused by the flux of secondary photons induce a spurious correlation with a minimum characteristic delay $\tau_{dd}$ equal to the light travel time delay between SPADs. As the full path is $\simeq$40 cm, $\tau_{dd} \simeq 1.3$ ns. This value is in agreement with the observed start of the rising edge of the peak (Figure~\ref{fig:g2taufilters}), while the decay has a characteristic time scale of 4-5 ns which depends on the intrinsic physical properties of the SPAD detectors. For this reason we decided to remove the intervals $\tau \simeq [-18, -1]$ ns and $\tau \simeq [1, 18]$ ns from the analysis (Figure~\ref{fig:g2taufilters}, panel B). We conservatively considered -18 ns and +18 ns as lower and upper bounds of these intervals to be sufficiently far away from the tails of the distribution of secondary photons.

After removing the intervals $\tau \simeq [-18, -1]$ ns and $\tau \simeq [1, 18]$ ns, the measurements of $g^{(2)}$ with the two filters have a similar pattern and show clearly a deep ($\sim$10\%) anti-correlation at $\tau \simeq 0$ delay with a characteristic width of $\sim$1 ns, and smaller oscillations and an overall loss of efficiency (of the order of 1\%) at large delays (Figure~\ref{fig:g2taufilters}, panel B). This problem was already identified in the preliminary measurements reported in \cite{2016SPIE.9907E..0NZ}. This behavior does not appear in the measurement of $g^{(2)}$ between the two telescopes and originates from spurious (anti-)correlations between the acquisition channels of the single front-end electronic board that reads the signals from all the detectors. These (anti-)correlations and oscillations of the electronics produce rather stable patterns in the measurement of $g^{(2)}$ over an entire night of observation. Despite these effects, Figure~\ref{fig:g2taufilters} (panel C) shows that a clear 'excess correlation' between -400 ps and 0 is present in the data acquired with the II filter compared to those acquired with the H$\alpha$+ND1 filter.
Indeed, this is what we are looking for, as the coherence of the photons acquired with the very narrow band II filter is approximately 10 times larger than that of the photons acquired with the H$\alpha$+ND1 filter (being the II filter width 10 times smaller)\footnote{The ND1 filter was inserted only for the purpose of limiting the rate to manageable values, and comparable to those of the measurements with the II filter.}.
Therefore, to remove these systematics and extract the actual signal, we subtracted the average values of $g^{(2)}$ measured with the H$\alpha$+ND1 filter (used for calibration) from those measured with the II filter (see eq.~[\ref{eq:eq3}]). The residual systematic error on $g^{(2)}$ after applying this procedure is typically $\lesssim$0.0005.

\section{Results}
\label{sect:results}

\subsection{Temporal correlation at zero baseline}
\label{sect:cor-zero}

\begin{figure}
	\includegraphics[angle=0, width=0.95\columnwidth]{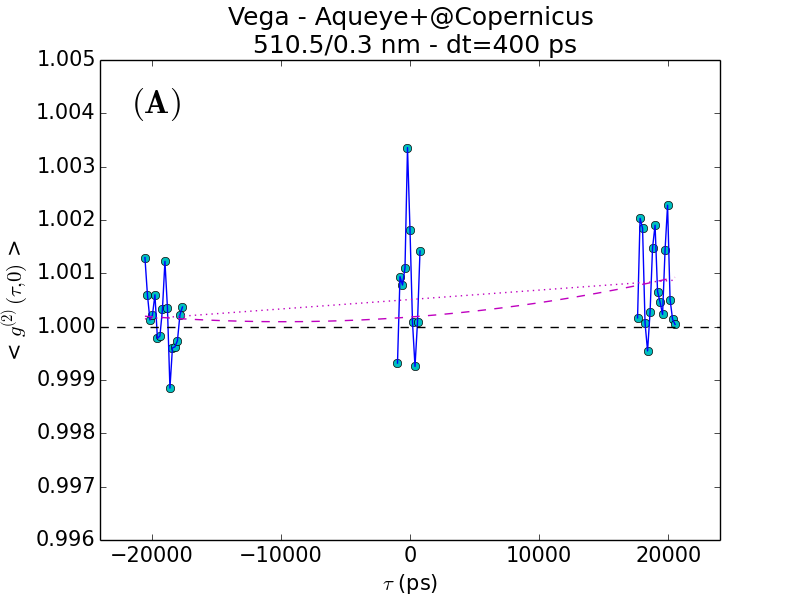}
	\includegraphics[angle=0, width=0.95\columnwidth]{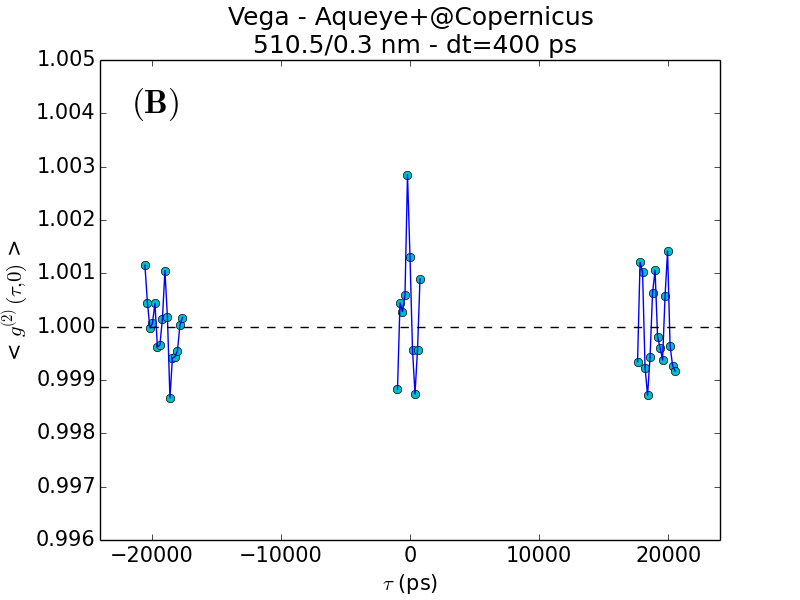}
    \caption{Calibrated temporal correlation at zero-baseline $< g^{(2)} (\tau,0) >$ for all the Aqueye+ observations of Vega reported in Table~\ref{tab:observations}. The time bin is $dt = 400$ ps. {\it Panel (A)}: $< g^{(2)} (\tau,0) >$ fitted with a first order polynomial $p(\tau)$ (excluding all the points around the peak; {\it magenta dashed line}) and a parabola (excluding only 5 points around the peak; {\it magenta dotted line}). {\it Panel (B)}: $< g^{(2)} (\tau,0) >$ after correcting with $1-p(\tau)$ (see text for details). }
    \label{fig:g2tau-1}
\end{figure}

\begin{figure}
	\includegraphics[angle=0, width=0.95\columnwidth]{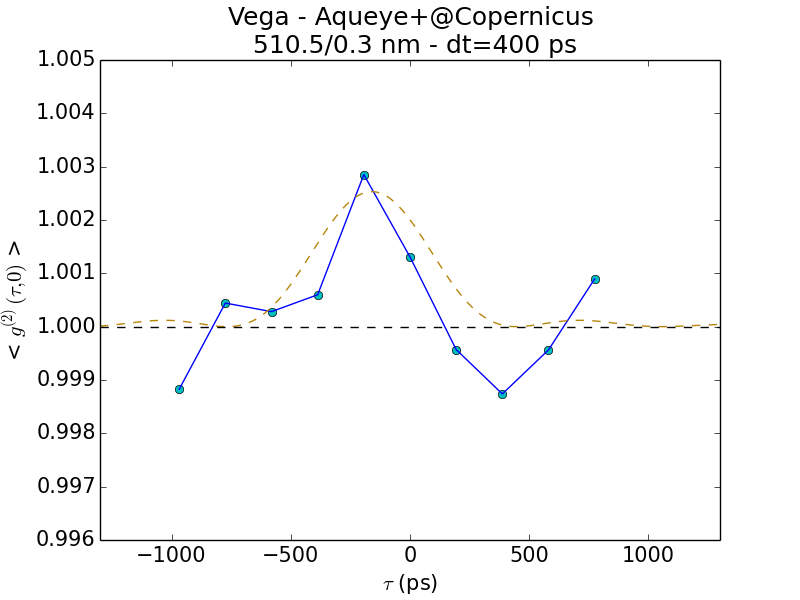}
    \caption{Calibrated temporal correlation at zero baseline $< g^{(2)} (\tau,0) >$ in the interval of delays $\tau = [-1,0.8]$ ns for all the Aqueye+ observations of Vega reported in Table~\ref{tab:observations}. The time bin is $dt = 400$ ps and the curve is corrected with $1 - p(\tau)$ (see text for details). The {\it dashed (yellow)} line shows the expectedected profile of the temporal correlation. }
    \label{fig:g2tau-2}
\end{figure}

For the measurement of the correlation at zero baseline, we considered only the observations performed with Aqueye+, that have significantly higher counting statistics.
As explained in the previous Section, we successfully removed some systematics and cross-talk effects subtracting the average value of $g^{(2)}$ measured with the H$\alpha$+ND1 filter from that measured with the II filter (eq.~[\ref{eq:eq3}]), and considering only the delay intervals $\tau = [-20.5, -18]$ ns, $\tau = [-1, 0.8]$ ns and $\tau = [18, 20.5]$ ns. The final calibrated value of $g^{(2)}$ for all the Aqueye+ observations of Vega is shown in Figure~\ref{fig:g2tau-1}. A peak in the degree of correlation at around zero delay is clearly visible. The value is: $<g^{(2)} (0,0)> = 1.0034$. At large delays $<g^{(2)} (\tau,0)>$ shows large random oscillations with root mean square (rms) $\sigma_{|\tau| \geq 18000 \, {\rm ps}} = 0.0008$. These fluctuations are dominated by the statistical uncertainty of the measurements, as they show the $t^{-1/2}$ decrement expected if the error is dominated by counting statistics. The estimated signal-to-noise ratio is then $S/N = (< g^{(2)} (0, 0 ) > - 1)/\sigma_{|\tau| \geq 18000 \, {\rm ps}} \simeq 4.2$.

A residual systematic offset of $< g^{(2)} (\tau, 0) >$ (that is larger at positive delays) is visible in the data and was fit with a first order polynomial $p(\tau)$ (excluding all the points around the peak). Subtracting $1-p(\tau)$, the value at the peak decreases (  $< g^{(2)} (\tau, 0) > = 1.0029$), as well as the rms at large delays ($\sigma_{|\tau| \geq 18000 \, {\rm ps}} = 0.0007$; Figure~\ref{fig:g2tau-1}, panel B). A parabolic fit including all but 5 points around the peak is equally acceptable and gives similar results. The origin of this small residual systematics has to do with different factors, such as residual calibration uncertainties, spectral dependences of the delay distributions of the secondary photons produced by the detectors, and/or rate-dependent effects (there are small differences in the average rates with the two filters). The systematic offset visible in Figure~\ref{fig:g2tau-1} may be considered as the ultimate limit for the accuracy achievable with our present instrumentation. After correcting for it (using different fitting functions and number of points in the fit), the estimated signal-to-noise ratio of the measurement is in the range $S/N \simeq 3.8-4.3$). This value is consistent with that calculated above and with the expected $S/N$ reported in equation~(\ref{eq:sn}), confirming that the measurement is significant.

Figure~\ref{fig:g2tau-2} shows an enlargement of $< g^{(2)} (\tau, 0) >$ around the peak (after subtracting $1-p(\tau)$), along with the expected profile of the temporal correlation (not fitted but simply overplotted, assuming an effective bandpass 25\% wider than the nominal FWHM of the II filter). The peak is clearly shifted by 160 ps towards negative delays (because of a residual difference of a few cm in the length of the cables connecting the detectors to the readout electronics), but the overall agreement is very good.

\subsection{Temporal correlation on a 4-km baseline}
\label{sect:cor-4km}

\begin{table}
\caption{Instrumental delays between IFI+Iqueye@Galileo and Aqueye+@Copernicus.}
\label{tab:delays}
\centering
\begin{tabular}{l c c}
\hline\hline
                                        &  Difference$^1$ (mm)  &   Delay$^1$ (ns)  \\
\hline
\hline
Equivalent focal lengths$^2$            &     -4215.5           &      -14        \\
Mirror distances$^3$                    &      1000             &        3        \\
IFI (instrument)                        &      1450             &        5        \\
IFI (optical fiber)                     &     20000             &       67        \\
Electric cables                         &    -12000             &      -40        \\
GPS antenna$^4$                                 &       --              &     --          \\
\hline
Total                                   &      6234.5           &       21        \\
\hline
\end{tabular}
\begin{minipage}{8.2 cm}
$^1$ IFI+Iqueye - Aqueye+ \\
$^2$ Copernicus telescope 16315.5 mm, Galileo telescope 12100 mm \\
$^3$ Referred to the intersection of the hour angle and declination axes \\
$^4$ Difference of the GPS antenna height relative to the intersection of the hour angle and declination axes
\end{minipage}
\end{table}

\begin{figure}
	\includegraphics[angle=0, width=0.95\columnwidth]{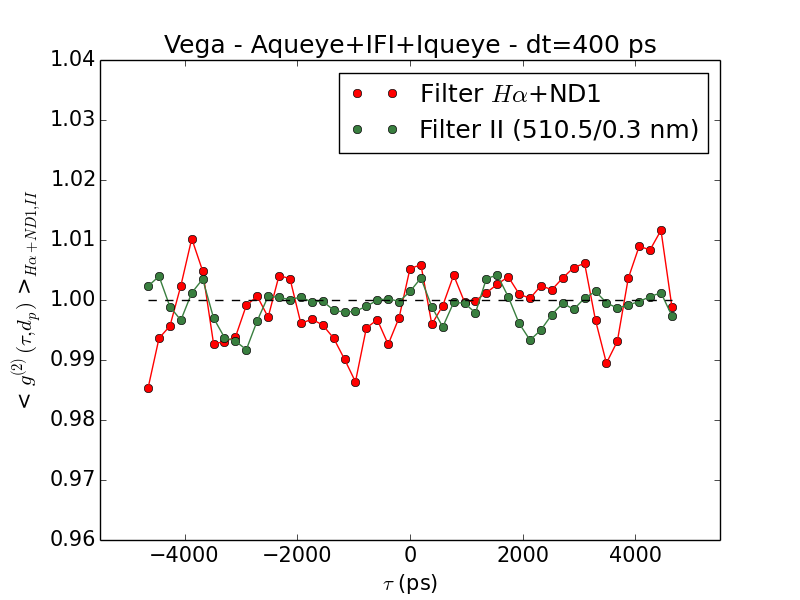}
    \caption{Temporal correlation on a $\sim$km baseline $< g^{(2)} (\tau,d_p) >_{\rm H\alpha+ND1,II}$ for all the Aqueye+IFI+Iqueye Nov 2019 observations of Vega reported in Table~\ref{tab:observations} and for the two adopted combinations of filters, H$\alpha$+ND1 and II. The time bin is $dt = 400$ ps. Data is corrected for the light travel time delay between telescopes. The projected baseline $d_p$ varied in the range 1535-2418 m. }
    \label{fig:g2taufilters_a+i}
\end{figure}

In Nov 2019 we performed simultaneous observations of Vega with both Aqueye+ at the Copernicus telescope and IFI+Iqueye at the Galileo telescope, forming an interferometer with a baseline of 1-4 km (Table~\ref{tab:coordinates}). To calculate the degree of coherence between the signals of the two telescopes they have to be properly corrected for the light travel time delays at the two sites. The relative delay is mostly caused by the light travel time distance projected along the direction of the star, with additional instrumental delays induced by differences in the focal lengths, position of the mirrors of the two telescopes, length of the electric cables, height of the GPS antenna (the GPS receiver is part of the acquisition and timing system of Aqueye+ and Iqueye, e.g. \citealt{2009JMOp...56..261B}) and, for Iqueye, by the additional optical path inside IFI and the optical fiber. The light travel time delay and the projected distance between the two telescopes are calculated as a function of the position of the star on the sky and of the telescope coordinates (Table~\ref{tab:coordinates}), while the instrumental delays are summarized in Table~\ref{tab:delays}. The total delay is continuously added to the photon arrival times of IFI+Iqueye before performing the correlation. The projected telescope separation during the Nov 25 and 28, 2019 observing nights was varying in the range 1589-2023 m and 1535-2418 m, respectively.

The average value of the discrete degree of coherence for all the simultaneous acquisitions obtained with the H$\alpha$+ND1 and the II filters is shown in Figure~\ref{fig:g2taufilters_a+i}. The adopted time bin is $dt \simeq 400$ ps and the average is performed over all the combinations of the instrumental sub-apertures. The calculation of $g^{(2)}$ was performed as described in Section~\ref{sect:analysis}. None of the systematic effects that affect the Aqueye+ measurements at zero baseline and discussed in the Appendices is visible in Figure~\ref{fig:g2taufilters_a+i}. Since in this case we used two independent front-end electronic boards and acquisition systems, it is clear that all systematic effects that appear at zero baseline originate from spurious (anti-)correlations between different channels intrinsic to a single front-end electronics.

For consistency with the approach adopted for zero baseline, we calibrated the measurements subtracting the average values of $g^{(2)}$ measured with the H$\alpha$+ND1 filter from those measured with the II filter (eq.~[\ref{eq:eq3}]). No peak in the degree of correlation at around zero delay is visible and the fluctuations of $g^{(2)}$ are dominated by statistical uncertainty. However, because of the variable observing conditions, during the Nov 2019 run the total simultaneous acquisition time (and the average count rate) with the H$\alpha$+ND1 filter was shorter than that with the II filter (28 minutes versus 40 minutes). Therefore, the random noise on the calibrated $< g^{(2)} (\tau, d_p) >$ (difference between the two filters) is significantly larger than that on the II measurements $< g^{(2)} (\tau, d_p) >_{II}$. As no systematic effect is visible in Figure~\ref{fig:g2taufilters_a+i}, we decided to use only the 'uncalibrated' measurements taken with the sole II filter to place a more stringent constraint on the absence of correlation. Given that no prominent peak is present in the II measurements in Figure~\ref{fig:g2taufilters_a+i} and assuming that they are randomly distributed in $\tau$, we consider them as representative of a series of measurements at zero delay and estimate $g^{(2)} (0, d_p)$ taking the average for all $\tau$. The resulting value of the degree of coherence for the simultaneous Aqueye+IFI+Iqueye measurements of Nov 2019 is: $< g^{(2)} (0, d_p) >_{A+I} = 0.999$. We estimated the uncertainty of the measurement from the standard deviation of $g^{(2)}$ at all delays, obtaining $\sigma_{A+I} = 0.003$. Despite the non-negligible uncertainty, our measurement is thus consistent with the absence of correlation, as expected for Vega on a projected baseline of $\sim$2 km (see below).

\subsection{Spatial correlation}
\label{sect:spat-cor}

\begin{figure}
	\includegraphics[angle=0, width=0.95\columnwidth]{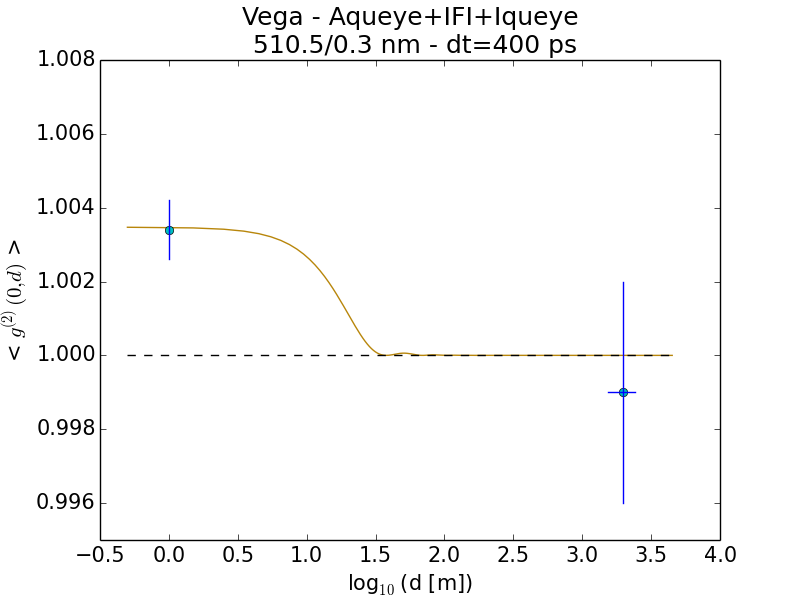}
    \caption{Spatial correlation $< g^{(2)} (0,d)>$ for the Aqueye+IFI+Iqueye 2019 observations of Vega. The yellow solid line represents the theoretical $g^{(2)}$ for a uniform brightness disc of 3.3 mas (angular size of Vega; \citealt{2001ApJ...559.1147C,2012ApJ...761L...3M}).
    }
    \label{fig:g2sep_a+i}
\end{figure}

Figure~\ref{fig:g2sep_a+i} shows the two measurements of the degree of coherence for Vega reported in the two previous subsections as a function of telescope separation. The zero baseline refers to the separation of the centroids of the mirror segments, which is approximately 1 m, while the long baseline corresponds to a telescope separation between 1535 m and 2418 m, varying with the star position on the sky. As it can be seen from Figure~\ref{fig:g2sep_a+i}, the measurements are fully consistent with the expected degree of spatial coherence for a source with the angular diameter of Vega ($3.28 \pm 0.01$ mas, \citealt{2001ApJ...559.1147C}; 3.324 mas, \citealt{2012ApJ...761L...3M}), with a positive detection at zero baseline and no detection at a comparable level on a $\sim$km baseline.



\section{Discussion and conclusions}
\label{sect:discussion}

Our detection of the temporal correlation of a star at zero baseline and the measurement on long baseline represents another proof of principle for stellar intensity interferometry. Unlike the original Hanbury Brown and Twiss experiment, that correlated in real-time the photon intensities measured at two telescopes, the measurement reported here is obtained for the first time counting photon coincidences in post-processing by means of a single photon software correlator and exploiting entirely the quantum properties of the light emitted from a star. Working in post processing has also the non-negligible advantage that the data reduction chain can be repeated more times (as, in fact, we did in Asiago), enabling the possibility to check for systematics, tune the parameters of the analysis, optimize the procedure, and increase the accuracy of the results. In principle, it could also enable the computation of correlations among three or more telescopes.

Unfortunately, the limited collecting areas of the Asiago telescopes are not suitable to perform measurements on weak targets, and the separation of the telescopes is not adequate to resolve sources on a mas scale. Nonetheless, the Asiago experiment allowed us to carry out a preparatory activity for potential implementations of stellar intensity interferometry on long-baseline arrays of Cherenkov telescopes. As a matter of fact, future Cherenkov installations will have both large collecting areas and a large number of baselines, suitable for performing SII measurements and image reconstruction with an unprecedented spatial resolution \citep{2006ApJ...649..399L,2012MNRAS.419..172N,2012MNRAS.424.1006N,2013APh....43..331D,2013MNRAS.430.3187R,2019BAAS...51c.275K}.

On the other hand, further progress needs to be made in order to set up a multi-baseline photon counting intensity interferometer on an array of Cherenkov telescopes capable of performing imaging at 10-100 microarcsecond scales. First of all, a relative photon timing accuracy among different telescopes of $\sim$1 ns is needed to correlate the signals over short time bins $dt$ or high sampling frequencies (see eq.~[\ref{eq:sn}]), thus keeping the observing time within reasonable limits ($\sim$hours). In Asiago we achieve this goal using independent acquisition and timing systems at the two telescopes, each made of a Rubidium clock disciplined with a GPS receiver, that allow us a synchronization with UTC with an accuracy of $\sim$1-2 ns \citep{2009JMOp...56..261B,2009A&A...508..531N}. Nowadays, alternative solutions based on synchronization signals distributed through ethernet networks are available, but our approach retains the required reliability and accuracy, and could still be competitive for very long (several km or more) baselines. More challenging is the effective utilization of narrow band filters. While this is a common problem also for other SII implementations, it is particularly constraining for a single photon counting approach like ours because photon rates must be limited to affordable values without reducing the signal (i.e. without significantly attenuating the photon flux). For the very small f/numbers (i.e. $\sim$f/1) of the Cherenkov telescopes, the angle of incidence of the rays coming on the interferometric filter from the outer portion of the mirror is very large (tens of degrees). Consequently, the transmitted wavelength of such rays is significantly smaller than that of those coming at normal incidence \citep{2019BAAS...51c.275K}. This broadening of the transmitted bandpass $\Delta \lambda$ for a given photon rate reduces the $S/N$ ratio of a measurement. To narrow the filter bandpass while maintaining a good transmission efficiency, an appropriate solution is installing a (removable) optical module at the telescope focal plane, suitably designed to reduce the angle of incidence.

A further aspect to consider for the implementation of SII on large area telescopes is handling the very high expected photon rates. The detectors must sustain more than $10^8$  events/s and the acquisition electronics must be capable of coping with very high data rates (up to a few Gbit/s). In this respect, a number of selected components (SiPM detectors, Time-to-Digital Converters or Digitizer Cards with Field Programmable Gate Arrays and data compression, Computers with fast motherboards) with the required performance are presently available on the market. In addition, significant storage space and computational power are needed for saving and processing the large amount of acquired data. A post-processing approach similar to that adopted in Asiago can be applied to Small-Size Cherenkov Telescopes (SSCTs) of the 4-meter class or to larger area telescopes for weaker targets. We estimate that a $\sim$1 hour observation at a maximum rate of $\approx$100 Mcounts/s will produce a few Terabytes of raw data at each telescope. While these are significant but manageable numbers, the requirement in terms of computational time is rather demanding. Scaling from the processing time required for the data of the Asiago experiment, a $\sim$1 hour measurement of $g^{(2)}$ performed at the maximum rate, sufficient to reach a $S/N \sim 5$ with a filter having a bandpass of several nanometers, will require 14 hours for 8 baselines on a machine with 2000 CPU cores. For the typical effective area of a SSCT, the maximum rate is reached for the brightest stars ($V \sim 0$), while for a Large Size Cherenkov Telescope (20-meter class) for a star with $V \sim 4$.

\begin{figure}
	\includegraphics[angle=0, width=0.95\columnwidth]{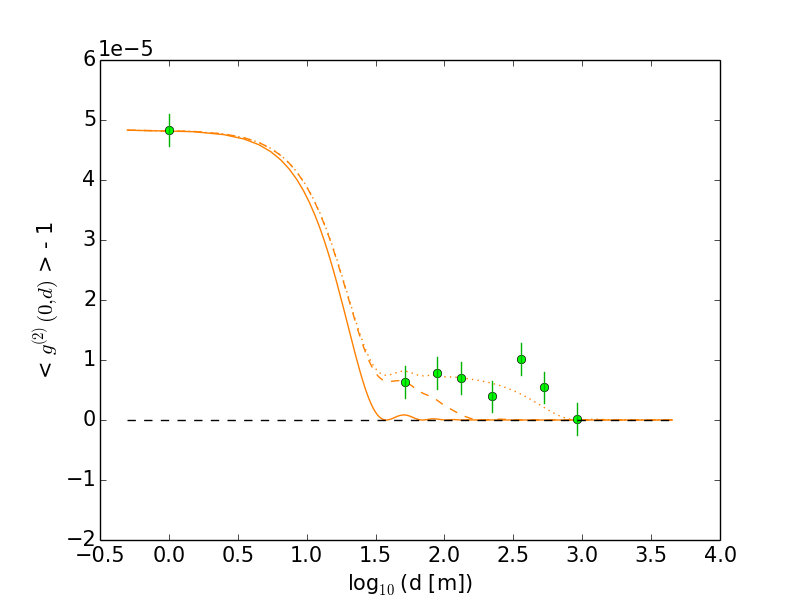}
    \caption{Simulated spatial correlation ($< g^{(2)} (0,d)> - 1$) for Cherenkov telescope observations. The orange solid line represents the theoretical $g^{(2)}$ for a uniform brightness disc of 3.3 mas emitting 70\% of the source photon flux. The other lines are the theoretical $g^{(2)}$ for a bright spot emitting 30\% of the source photon flux, overimposed on the 3.3 mas disc. Spot size: 130 $\mu$as ({\it orange dotted line}), 430 $\mu$as ({\it orange dashed line}).}
    \label{fig:g2sep_ssct}
\end{figure}

To illustrate the potential of the technique with the upgrades outlined above and with the resolving capabilities of a km baseline Cherenkov telescope array, we consider the hypothesis of a hot spot on the surface of a star like Vega and intensity interferometry observations carried out with the SSCTs. A detailed simulation of surface features reconstruction for an array of Cherenkov telescopes was already presented by \cite{2012MNRAS.424.1006N}. Here we describe only a specific example, providing details for a photon counting implementation within the framework discussed above. Figure~\ref{fig:g2sep_ssct} shows a simulated measurement at zero baseline plus 7 additional measurements on projected baselines from $\sim$100 m up to $\sim$1 km, assuming a time resolution of $\sim$1 ns, a bandpass of $\sim$5 nm and a count rate of $\sim$100 Mcounts/s. The total acquisition time per measurement is 4 hours. Simulated data are drawn from the expected theoretical value of $g^{(2)}$ for a bright spot of 130 $\mu$as emitting 30\% of the source photon flux, overimposed on a disc of 3.3 mas emitting the remaining 70\%. The best match is obtained for the same theoretical curve (reduced $\chi^2 = 1.0$ for 8 degrees of freedom). The other curves are not consistent with the simulated data (reduced $\chi^2 > 3.7$ for 8 degrees of freedom). We emphasize the importance of having a SII implementation capable of at least one simultaneous measurement at zero baseline, that permits to calibrate the contribution of the stellar component and reduce the uncertainty on the parameters estimation. The simulated measurements are consistent with the presence of a hot spot with a size 25 times smaller than that of the star. For thermal emission, the temperature of the spot would be significantly higher than that of the star. While the optical-UV spectrum would show evidence for such an additional hot component, its actual morphology and structure could only be investigated through interferometric observations. A km baseline interferometric observation has thus the potential to place a direct constraint on surface features as small as tens of $\mu$as on the surface of a star, and hence to probe magnetic phenomena, such as those inferred from the observation of rotation modulations and flaring activity in the Kepler light curves of numerous A-type stars \citep{2017MNRAS.467.1830B,2017ApJS..232...26V}. Larger effective collecting areas such as those obtained combining Small Size with Medium/Large Size Cherenkov Telescopes would provide the required photon flux even for weaker targets or lower-contrast spots/features. With SII on km projected baselines one is then moving into novel and previously unexplored parameter domains in stellar Astrophysics, with the possibility to achieve imaging capabilities and angular resolutions in the optical band close to those attained at mm wavelengths with the Event Horizon Telescope.

\section*{Acknowledgements}

We thank the referee, Theo Ten Brummelaar, for his useful and constructive comments. We would like to thank Enrico Verroi, Mauro Barbieri, Paolo Ochner, Gabriele Umbriaco, Luigi Lessio, Giancarlo Farisato, Paolo Favazza and all the technical staff at the Asiago Cima Ekar and Pennar Observatories for their valuable help and operational support. We acknowledge financial contribution from Fondazione Banca Popolare di Marostica-Volksbank. Based on observations collected at the Copernicus telescope (Asiago, Italy) of the INAF-Osservatorio Astronomico di Padova and at the Galileo telescope (Asiago, Italy) of the University of Padova. This research made use also of the following PYTHON packages: MATPLOTLIB \citep{2007CSE.....9...90H}, NUMPY \citep{2011CSE....13b..22V}.

\section*{Data Availability}

Original event lists are stored in the Aqueye+Iqueye Public Data Archive, reachable from the Aqueye+Iqueye project page: https://web.oapd.inaf.it/zampieri/aqueye-iqueye/ and are available upon request. All analyzed data needed to evaluate the conclusions in the paper are present in the paper and/or the Appendices.




\bsp	
\label{lastpage}
\end{document}